\documentclass[]{aastex63} 

\shorttitle{Inverse Raman scattering and the diffuse interstellar bands}
\shortauthors{F. Zagury}
\begin{document}

\title{
Inverse Raman scattering and the diffuse interstellar bands: an exploration of the systemic interconnections between spontaneous and inverse Raman scattering and extended red emission, Red Rectangle bands, and diffuse interstellar bands}

\author[0000-0002-0786-7307]{Fr\'ed\'eric Zagury}
\affiliation{Fondation Louis de Broglie \\
23 rue Marsoulan \\
75012 Paris, France}
\nocollaboration{1}

\begin{abstract}
First identified in 1964, inverse Raman scattering (IRS) is a nonlinear stimulated phenomenon that induces Raman scattered absorptions where Raman emissions would be expected.
While IRS is less well-known than stimulated Raman scattering (SRS) and coherent anti--Stokes Raman scattering (CARS), this study highlights its significance in analyzing the spectra of stars  located  in the distant background of  HI interstellar clouds.
Specifically, ultraviolet emission lines Raman scattered by atomic hydrogen, typically observed in emission at wide scattering angles in the optical spectra of symbiotic stars and nebulae, should appear as IRS absorption features in the optical spectra of the background stars.
I show that  all known interstellar Raman scattered emission lines in the  H$\alpha$ wavelength region are detected  in absorption as diffuse interstellar bands (DIBs) in the spectra of reddened stars, and conclude that IRS by atomic hydrogen resolves   the longstanding puzzle of the processes involved in producing these bands, and perhaps also explains the equally mysterious 2200~\AA\ bump of ultraviolet extinction curves.
This identification of  DIBs as IRS HI absorptions sheds new light on the perplexing relationship between DIBs and the Red Rectangle nebula  emission bands (RRBs).\\
The conditions under which DIBs are detected highlight the importance of considering the physical relationship between the observer, the HI medium, and the direction of the illuminating radiation field (i.e., the geometry of the observation) in observations of HI interstellar matter. 
Observing in the direction of the radiation field or on its side determines whether IRS, yielding DIBs and the 2200~\AA\ bump, or spontaneous Raman scattering at wide scattering angles, resulting in ERE, Raman scattered emission lines (including RRBs), and the unidentified infrared bands, will be observed.
\end{abstract}

\section{Introduction:  the  significance of the Jones and Sto\"icheff  article for astrophysics}\label{intro}
The advent of lasers in laboratories in 1960  revolutionized research in nonlinear optics \citep[][]{schawlow63}. 
\citet{eckhardt62}  successfully stimulated Stokes Raman scattering in Raman--active liquids, generating laser--like coherent Raman scattered waves with intensities orders of magnitude greater than those of spontaneous Raman scattering (sRS). 
This discovery was soon complemented by the observations of  \citet{minck63} and \citet{stoicheff63} of stimulated Raman scattering at  both Stokes and anti--Stokes frequencies.
These findings enabled the development of two major coherent Raman scattering (CRS) techniques used in various fields of physics and medicine: stimulated Raman scattering \citep[SRS,][]{prince17} and coherent anti-Stokes Raman scattering  \citep[CARS,][]{polli18}.
The practical applications of SRS and CARS have somewhat overshadowed a third nonlinear stimulated Raman process, inverse Raman scattering (IRS), which  Jones and Sto\"icheff described in a concise 1964 article \citep[][]{jones64}. 

In this paper I show that IRS emerges as a major astrophysical phenomenon when the line of sight intersects an interstellar cloud and a star located in the cloud's distant background, and that this phenomenon provides a compelling explanation for the diffuse interstellar bands (DIBs).
This result complements a previous study which established that prominent spectral emission features associated with the  optical/infrared reflected light from nebulae, such as the extended red emission (ERE) and the unidentified infrared bands (UIBs), tend to cluster around hydrogen's Balmer and infrared lines \citep{apj23}.
That study further highlighted a  remarkable quantitative concordance between observational data and theoretical predictions, leading to the conclusion that ERE results from sRS by atomic hydrogen of the continuum near Ly$\beta$.

The experimental arrangement used by \citet[][]{jones64} to demonstrate inverse Raman scattering (IRS) resembles an SRS experiment (Section~\ref{irs}).
However, the IRS experiment diverges from SRS in that the intensity of the incident (pumped) laser is maintained below the SRS threshold, and the stimulating laser at the output Raman frequency exhibits a broad spectral width akin to a continuum  \citep[see also][]{polli18}.
Under these conditions, at the Raman frequency observers detect  a prominent absorption along the precise alignment of the laser and the continuum, as opposed to the anticipated Raman scattered emission line in an SRS experiment.
 In the conclusion of their groundbreaking work, Jones \& Sto\"icheff note the potential significance of IRS for astrophysics.

In 1989, \citet{schmid89} resolved a longstanding problem in the spectroscopy of symbiotic stars by identifying an ultraviolet doublet of oxygen Raman scattered by atomic hydrogen in the visible spectrum. 
Schmid's finding, supported by a suggestion in \citet{nuss89} that  Raman scattering could be a common astrophysical process and justify optical emission lines yet  to be identified\footnote{"We show that Raman scattering deserves to be included in the diagnostic tools of the spectroscopy of gaseous nebulae and emission regions of galactic nuclei. We point out that Raman scattering may be the source of some up to now unidentified emission lines." \citep[][p.~L27]{nuss89}.},  initiated a new area of research.
Over the past three decades, attention has focused on the vicinity of hydrogen Balmer lines, especially  H$\alpha$, where the probability of detecting hydrogen Raman scattering is optimized (Section~\ref{srs}).
A dozen Raman scattered ultraviolet emission lines by hydrogen from atoms such as oxygen, helium, and calcium, most of which lie close to H$\alpha$, have so far been highlighted in symbiotic stars and nebulae.
The mechanism for interstellar HI Raman scattering  involves the excitation of atoms in the HII gas, which shields the HI gas from radiations above the Lyman limit, followed by sRS of the resultant ultraviolet emission lines within the HI region and toward the observer (thus at large scattering angles).

Jones \& Stoicheff did not specify what  applications of IRS to astrophysics they envisioned.
However,  it is evident that as the scattering angle approaches zero, such as when a star is observed through an HI cloud, the observational setup mirrors their IRS experiment (Sections~\ref{irs}--\ref{irsast}).
Not only does the star excite ultraviolet transitions at the outskirts of the HI cloud, but it also provides the requisite stimulating optical continuum for IRS by  atomic hydrogen. 
If the star is situated  in the far background of the interstellar cloud and thereby ensures a coherent radiation field over the largest possible area at the cloud's position, optimizing  IRS conditions (Section~\ref{irsast}), the same Raman scattered lines seen in emission in the reflected starlight of symbiotic stars and nebulae should manifest in absorption in the star's optical spectrum.

In Sections~\ref{intirs}--\ref{ds}, I show that DIBs (Section~\ref{dibs}) and potentially also the 2200~\AA\ bump of ultraviolet extinction curves,  can be attributed to HI IRS.
All identified interstellar Raman lines within the H$\alpha$ spectral region  align with entries in DIB compilations (Section~\ref{dibid}).
DIB wavelengths correspond to  the Raman scattered position of the source ultraviolet emission lines' central wavelengths, which is slightly different from the central wavelengths of the Raman scattered lines (Section~\ref{glin}).

Section~\ref{rrbs} examines the interrelations between DIBs, ERE, and  the Red Rectangle emission bands  (RRBs) superimposed on ERE \citep[e.g.,][]{schmidt80,scarrott92,vanwinckel02,witt20}.
The hypothesis  suggesting that prominent RRBs are associated with some of the deepest DIBs was questioned by \citet{glinski02} due to   a slight redshift of RRBs relative to their corresponding DIBs (Sections~\ref{ass}--\ref{vw}).
However, this redshift parallels the shift between Raman emission lines and DIBs (as mentioned above), and supports the Raman scattering origin of RRBs as evidenced by their broad asymmetric profiles \citep{kokubo24}.
Section~\ref{rrbraman} concludes that ERE is Raman scattering of the continuum near Ly$\beta$ and that RRBs are HI Raman scattered emission lines; DIBs associated with RRBs are their IRS counterparts observed in the spectra of reddened stars.
The structure of the DIB spectrum (Section~\ref{ds}), concentrated in the ERE wavelength domain and to a lesser extent around hydrogen's optical and infrared lines, confirms these conclusions and explains the improbable superposition of  DIB spectral density (number of DIBs per unit wavelength interval) and  ERE profiles \citep[Figure~1 in][]{witt14}.
Possible applications of the nonlinear character of IRS are examined in Section~\ref{bump}.

As explained in Sections~\ref{rrbs} and \ref{ds}, the connection of DIBs to ERE and RRBs  is due to the shared nature of IRS and sRS as Raman phenomena.
However,  forward--directed IRS and isotropic sRS are observed under widely different conditions that affect whether DIBs or ERE and RRBs will be detected (Section~\ref{geo}).
Due to its low cross--section outside Lyman wavelengths and to its isotropy  (Section~\ref{srs}) sRS by atomic hydrogen in its ground state requires intense ultraviolet radiation.
It is observed in the form of  ERE and RRBs within the optical spectrum\footnote{and likely also blue luminescence \citep{vijh05}  and the unidentified infrared bands (UIBs) \citep[][]{apj23}.} in the reflected light at large scattering angles of bright stars illuminating nearby interstellar matter (such as in symbiotic stars and nebulae).
Nonlinear stimulated  IRS  manifests only when  the line of sight intersects HI interstellar matter against a background radiation field (e.g., a star).
Under these conditions,  DIBs in the optical spectrum and the 2200~\AA\ bump of ultraviolet extinction curves  (Section~\ref{bump}) are observed.
The geometry of an observation, that is the geometrical relationship between the observer, the scattering medium, and the radiation field, thus emerges as  a critical parameter in the analysis of spectral observations of  photodissociation fronts (Section~\ref{geo}).
Section~\ref{dis} summarizes the paper's findings and discusses their implications.
\section{Essentials on hydrogen Raman scattering  and Inverse Raman Scattering} \label{rs}
\subsection{Spontaneous Raman scattering by atomic hydrogen} \label{srs}
In HI interstellar clouds, photons undergoing Raman scattering\footnote{Raman scattering by hydrogen originally on level $n_0$  and left on level $n_1$ will be denoted as $\Gamma_{n_0\rightarrow n_1}$, and inverse Raman scattering as $\Gamma^i_{n_0\rightarrow n_1}$. A detailed analysis of cross--sections of $\Gamma_{1\rightarrow n}$ for $n=2$ to 4, emphasizing the importance of considering hydrogen's sub--levels, can be found in the recent \citet[][]{kokubo24}. See, in particular, Figure 3 in his paper.} by hydrogen at rest ($\Gamma_{1\rightarrow n}$) have energies between Ly$\alpha$ and Ly$_\infty$.
They are degraded into optical and infrared photons because of  the exceptionally large $\Gamma_{1\rightarrow n}$ Raman shifts.
The cross-section for $\Gamma_{1\rightarrow n}$ goes through asymptotic peaks at Lyman wavelengths Ly$_p$ with ordinal values of $p>n$ and troughs in between, as discussed in \citet{apj23}.
Consequently, Raman scattered emission lines by hydrogen tend to aggregate near the optical and infrared Balmer, Paschen, and other lines of the following hydrogen series.

There is a sharp contrast between  $\Gamma_{1\rightarrow 2}$ scattering  of an atomic emission line proximal to Ly$\beta$, resulting in a solitary emission near H$\alpha$, and the  $\Gamma_{1\rightarrow 2}$  scattering of an emission line proximate to Ly$_p$ with $p>3$, which generates $p-3$ $\Gamma_{1\rightarrow n}$ ($n<p$) Raman scattering possibilities.
Specifically, photons close to Ly$_p$ with $p>3$  may be degraded through  $\Gamma_{1\rightarrow 2}$ into Balmer photons close to H$_{p-1}$, but they can also undergo Raman scattering through any of the $p-3$ $\Gamma_{1\rightarrow p-1}$, $\Gamma_{1\rightarrow p-2}$,..., $\Gamma_{1\rightarrow 3}$ processes.
The Raman scattered photons are subsequently distributed between the vicinities of hydrogen resonances in the far-red (above Pa$_\infty$ at 8200~\AA) and infrared regions of the spectrum.
By contrast, ultraviolet photons near Ly$\beta$ ($p=3$) undergo Raman scattering  only via $\Gamma_{1\rightarrow 2}$ and end up near H$\alpha$. 

Consequently, the H$\alpha$ region at large (between H$\beta$ and Pa$_\infty$) should manifest the most intense Raman scattered ultraviolet transitions from atoms and molecules of the interstellar medium.
These emissions must also diminish in intensity on each side of H$\alpha$ and of the H$\alpha$ region.
Additionally, due to the rapid narrowing of  the  [Ly$_p$, Ly$_{p-1}$] interval  with increasing $p$, the density distribution of detected Raman scattered lines should  peak at H$\alpha$  and diminish on both sides.

sRS by atomic hydrogen at rest is, like Rayleigh scattering, a quasi isotropic process, and its cross--section decreases steeply outside Lyman wavelengths.
Consequently,  detection of sRS in space necessitates intense, ultraviolet--dominated radiation fields, typically found near hot stars.
These conditions increase the intensity of sRS and heighten its contrast with the optical background.
Consequently, Galactic observations of HI sRS have been confined to symbiotic stars and nebulae.
ERE is barely observed for radiation fields with temperatures below $10^4$~K \citep{darbon99}.

\subsection{Stimulated and inverse  Raman scattering} \label{irs}
\citet{jones64} derived the principles of stimulated Raman emission (SRE, including Stokes and anti-Stokes scattering) and inverse Raman scattering from a common equation.
This equation involves  a ''pumped'' laser at exciting frequency $\omega_0$, and a stimulating laser (frequency $\omega_s=\omega_0\pm \omega_r$) that triggers Raman scattering in a medium of particles with intrinsic frequency $\omega_r$.
When the pumped laser operates below the SRE  threshold and  the stimulating radiation has a sufficiently broad spectral width (wide enough to be locally treated as a continuum),  Raman lines typically  observed as emissions in SRE appear as absorptions within the continuum after the lasers have traversed the medium \citep{jones64,polli18}.

IRS has no threshold, although its intensity can be heightened near the SRE threshold  \citep[][]{dumartin65a,dumartin65b,dumartin67a,dumartin67b}. 
A nonlinear phenomenon, IRS depends on the intensity of both the pumped and stimulating lasers and can also generate new frequencies.
Frequency coupling and generation of harmonics at frequencies  $\omega_0\pm n\omega_r$ ($n$ an integer), summation and difference of frequencies when particles exhibit multiple resonances  \citep[][]{stoicheff63,minck63,chiao64,maker65} were also  observed in IRS experiments \citep{jones64,dumartin67a,mitrofanov21}. 
Contrary to nearly isotropic sRS, IRS is forward directed.
\subsection{Inverse  Raman scattering in astronomical conditions} \label{irsast}
Stars  (preferably of A--O spectral type), the average background starlight (see Section~\ref{o1}), galaxies, are common sources of ultraviolet light that can excite interstellar foreground atomic hydrogen and  also provide an optical continuum.
To observe IRS, the celestial source must be situated behind the interstellar HI medium on the observer's line of sight.
The pumped wavelength can be an emission line, most often close to a Lyman resonance (Ly$\beta$ especially, Section~\ref{srs}),  carried by the source or excited by  absorption of its ultraviolet light by hot gas at the edge of interstellar clouds \citep[for the OVI ultraviolet doublet, see][and references therein]{cowie79,bowen08}.

As per classical wave optics, a star disturbance at wavelength $\lambda$ measured at the observer position is the summation of the disturbances issued from half the first Fresnel zone (viewed by the observer) at the intervening HI cloud location \citep{an17}.
The wider the first Fresnel zone, the more starlight will be affected  by HI IRS.
For IRS absorptions to manifest, the first Fresnel zone must be maximized and the star be far behind the HI cloud. 
\citet[][]{vdh} further clarifies\footnote{''
\emph{The amplitude at distance L beyond a plane wave front is such as if an area $l\lambda$ contributes with equal phase and the remaining part of the wave front not at all}''. \citet[][p.~21]{vdh}.
}  that the part of the scattered plane wave that determines its disturbance at the observer position is the first Fresnel zone (at the cloud position, viewed  by the observer).
Consequently, IRS absorptions in the star's spectrum will be detectable if starlight at the cloud position is coherent over an area larger than this first Fresnel zone, that is if the star is at least as far behind the cloud as the observer--cloud distance.
\begin{figure}
\resizebox{0.5\columnwidth}{!}{\includegraphics{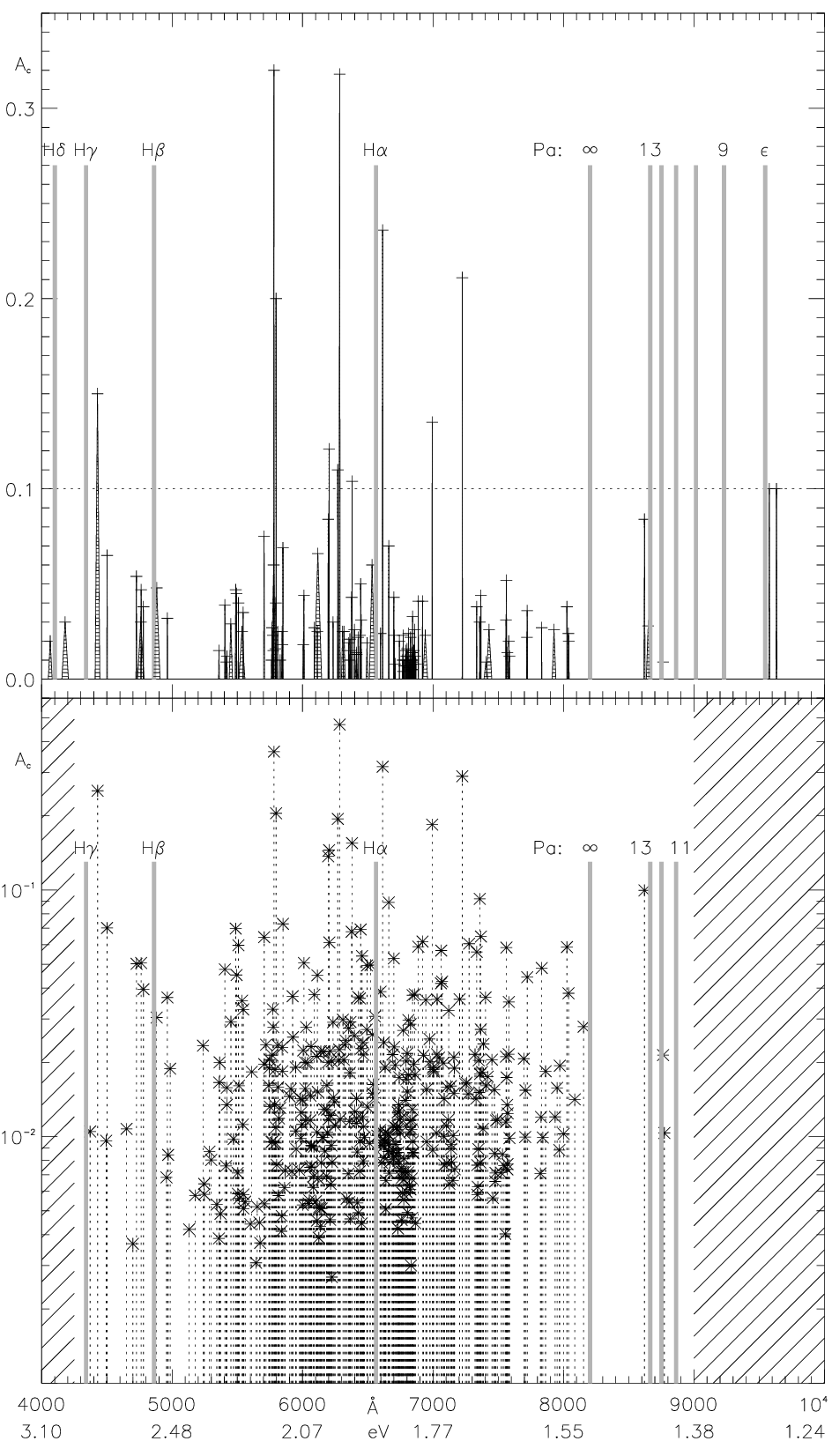}} 
\caption{DIB spectrum of HD183143 in 1995 and 2009 with hydrogen lines added (thick grey verticals).
Top plot: DIB spectrum ($\sim$100 DIBs) of HD183143 according to \citet[][his Figure~2]{herbig95}.
Each DIB is represented by a triangle of height the DIB's depth $A_c$ and basis twice the FWHM.
Twelve DIBs have $A_c>10\%$.
The few DIBs outside the H$\alpha$/ERE region  cluster around hydrogen's lines.
Bottom plot: DIB spectrum  of HD183143 derived from the catalog compiled by \citet[][]{hobbs09}.
I have estimated $A_c$ from their equivalent width (EW) and FWHM, as given in the catalog ($A_c\sim$ EW/FWHM).
Increased sensitivity essentially increases the number of DIBs around H$\alpha$.
} 
\label{fig:fig1}
\end{figure}
\section{Interstellar inverse Raman scattering} \label{intirs}
\subsection{DIffuse interstellar bands} \label{dibs}
Diffuse interstellar bands (DIBs) are enigmatic absorption features occurring at  fixed wavelengths in the spectra of stars (mainly in the optical spectrum, Figure~\ref{fig:fig1}) obscured by interstellar clouds in their far foreground \citep[no DIBs in circumstellar matter,][]{snow11,an17}. 
Observations suggest that DIBs arise from the outer region of interstellar clouds \citep[the skin effect, see][]{snow74,herbig95,snow02}.

Over the past three decades, advances in observational sensitivity have enabled a significant increase in detection of the number of DIBs. 
Current compilations include over 500 DIBs, in contrast to approximately a hundred cataloged in the 1990s and a mere 39 DIBs uncovered in 1975  \citep[][]{herbig75,herbig95,fan19}.
Most DIBs  appear as minor dips that account for at most a small percent of the continuum of a reddened star's spectrum   (bottom plot of Figure~\ref{fig:fig1}).
However, a few DIBs are much deeper, and twelve prominent DIBs have depths ($A_c$) exceeding 10\% of the continuum  (top plot of Figure~\ref{fig:fig1}).
The two deepest  DIBs ($\lambda\lambda$6284, 5780) in the spectrum of HD183143 reach $A_c\sim 30$\%.
The full width at half maximum (FWHM) of DIBs varies widely, ranging from less than 1~\AA\ to over 20~\AA. 
\subsection{DIB identifications} \label{dibid}
\subsubsection{The OVI doublet} \label{o6}
The first  identification \citep{schmid89} of sRS from HI interstellar gas was the scattered OVI ultraviolet doublet close to $\lambda\lambda$1031.9\footnote{In this paper, ultraviolet wavelengths are all vacuum wavelengths; optical wavelengths are air wavelengths, unless otherwise specified.},  1037.6~\AA\ observed in symbiotic stars.
This identification is remarkable since the theoretical Raman scattered wavelengths calculated by Schmid, based on assumed central wavelengths of 1031.923 and 1037.618~\AA\ for the source ultraviolet lines (and a Raman shift of 82258.95~cm$^{-1}$), fall at 6825.44 and 7082.40~\AA\ (6827.32 and 7084.35~\AA\ vacuum wavelengths), considerably blueward  from the observed wavelengths of  6829.5$\pm0.5$ and 7088$\pm1.5$ derived by \citet{allen80} from a large set of symbiotic stars.
As reported by Allen, the ratio of observed  line intensities' at peak wavelengths  is around 4.
The OVI Raman scattered doublet was also recognized in the spectrum of PNe such as NGC7027 and NGC6302 \citep{groves02,zhang05}.

The National Institute of Standards and Technology (NIST) database\footnote{Available at:  https://physics.nist.gov/asd.} provides slightly shorter wavelengths, 1031.912 and 1037.613~\AA, than Schmid for the two OVI lines, and a Raman shift of 82258.2~cm$^{-1}$.
With these adjusted values, the lines are Raman scattered at 6827.44 and 7083.80~\AA\ (6829.32 and 7085.75~\AA\ vacuum wavelengths), in better agreement with observations of symbiotic stars but still approximately 2~\AA\ bluer.

\citet{jenniskens94} reported a DIB at 6827.28~\AA\ in the spectrum of HD183143 that corresponds precisely to the strongest line of the Raman scattered OVI doublet ($\lambda$6827.44).
For the same star,  \citet{herbig95} measured  a depth $A_c$ of the order of 2.2\% of the continuum.
 The more sensitive  \citet{hobbs09}\footnote{See the full spectrum in the paper's online material at: https://iopscience.iop.org/article/10.1088/0004-637X/705/1/32\#apj317757f11  (tarball of figure set images).} spectrum of HD183143 indicates a somewhat higher $A_c\sim2.7\%$. 
 
The weakest OVI Raman scattered line at 7083.80~\AA\   matches a DIB  reported thirteen years after DIB~6827.28  in the spectrum of HD204827 \citep{hobbs08}.
In the \citet{hobbs09} spectrum of  HD183143, the line is given at  7083.8~\AA\ (air wavelength), at the exact expected position of the Raman scattered line deduced from NIST's  OVI transition.
The $A_c$ ratio of the two DIBs is about 4 (from Hobbs et al.'s spectrum), similar to that of the OVI scattered doublet in symbiotic stars.
\subsubsection{The OI doublet} \label{o1}
 \citet{henney21} analyzed high resolution spectra of the complex inner Orion nebula within 2' from the strong  $\theta^1$ and $\theta^2$ Ori stars.
Spectra were taken in so-called Dark Bay and Orion Bar towards the edges of the field, and in the SW minibar and other sub--regions much closer to the field's central stars.
The data spectral coverage runs from 4595 to 9336~\AA.

Major DIBs~$\lambda\lambda$5780, 5797, 6284, 6614 are clearly present in the spectra, albeit at  much lower levels than typically observed in the spectra of O--B stars.
For example, in Henney's observation of the Dark Bay (see his Figure~7), DIB~5780 is a mere 8\% of the continuum, whereas it accounts for $A_c\sim  32\%$ of the continuum in HD183143 \citep[Figure~\ref{fig:fig1} and][]{herbig95}.
These four DIBs behave similarly across the field, being notably stronger in the densest and farthest regions from the Ori stars \citep[Figures~3 and 7 in][]{henney21}.
They thus appear to be unaffected by variations of the central stars' radiation field but increase with column density.
Given that DIBs are consistently observed as absorptions in background starlight, the observed Orion DIBs in Henney's field must originate from background starlight in the direction of the observer (which remains constant across the limited extent of the field) and are not associated with the Ori stars radiation field.

On the H$\alpha$ wings spanning the 6300--6800~\AA\ wavelength range, \citet[][his Section~2.2 and Figure~3]{henney21} identifies  two absorptions at  about 6633 and 6664~\AA\ (air wavelengths), which he attributes to  the OI doublet  $\lambda\lambda$1027.43, 1028.16~\AA\ Raman ($\Gamma_{1\rightarrow 2}$) scattered at air wavelengths 6633.3 and 6663.75~\AA\ (6635.2 and 6665.6~\AA\ vacuum wavelengths).
Unlike  the DIBs, these two features exhibit much greater strength in the SW minibar near the Ori stars compared to the Dark Bay, demonstrating their dependence on the Ori stars' radiation field.
As proposed by Henney, they likely result from spontaneous Raman scattering by HI of the doublet, in absorption, under illumination from the   ultraviolet radiation of these stars.

The OI Raman scattered doublet 6633, 6664 also coincides with the position of two DIBs.
The strongest (DIB~6633) was identified by  \citet{jenniskens94} in the spectrum of HD183143.
The second DIB was not resolved until observed in the spectra of HD204827 and HD183143 by \citet{hobbs08,hobbs09}.
These DIBs are recorded with central air wavelengths 6633.12 and 6663.71~\AA\ in the spectrum of HD204827 \citep{hobbs08}, and 6632.93 and 6664.05~\AA\ in HD183143 \citep{hobbs09}.
Both lines are blended with other DIBs, but their $A_c$ ratio should fall in the range 2--3 (see Hobbs et al.'s online material), which is consistent   with the ratio of the  $\Gamma_{1\rightarrow 2}$ scattered OI absorptions shown in Henney's Figure~3 (top plot).

The two DIBs can therefore be identified as $\Gamma_{1\rightarrow 2}^i$ IRS of the OI $\lambda\lambda$1027.43, 1028.16~\AA\ doublet.
Considerably less intense compared to the prominent $\lambda\lambda$5781, 5797, 6284, 6614 DIBs,  they are unlikely to be detectable in the background starlight of the Dark Bay observations, as outlined in \citet{henney21}.
In Henney's observations, the lines at 6633, 6664~\AA\ originate from $\Gamma_{1\rightarrow 2}$ Raman scattering by atomic hydrogen  at large scattering angles (see Section~\ref{srs}) of absorptions  in the Orion stars' radiation field.
ERE, also attributed to $\Gamma_{1\rightarrow 2}$ Raman scattering  \citep[of the continuum near Ly$\beta$,][]{apj23}, has been observed throughout the Orion nebula and, like the OI Raman scattered lines, decreases steeply with distance from the $\theta$~Ori stars \citep[Figure~5 in][]{perrin92}.
\subsubsection{HeII lines} \label{he}
HeII and HI have close ultraviolet transitions, facilitating  Raman scattering of  HeII lines by HI.
Ultraviolet HeII series neighboring Ly$\beta$, Ly$\gamma$, Ly$\delta$ are $\Gamma_{1\rightarrow 2}$ scattered near H$\alpha$, H$\beta$, and H$\gamma$ respectively \citep{lee06,vangro93}.
The strongest of these scattered emissions are expected   (Section~\ref{rs}) to be the near--H$\alpha$ lines close to 6545~\AA, followed by the near--H$\beta$ and near--H$\gamma$  lines at air wavelengths 4850 and 4331~\AA. 

Shortly after the papers by \citet{schmid89,nuss89},  \citet{vangro93} reported the 4331 and 4850~\AA\ lines in the spectrum of symbiotic RR~Tel.
As for the OVI lines, the observed central wavelengths of the Raman scattered HeII lines are slightly redshifted by 1--1.5~\AA\ from the scattered central wavelengths of the ultraviolet lines.
Although faint, the two lines are commonly detected in high ionization PNe possessing a neutral shell, including NGC7027 \citep[][]{pequignot03}.

The strongest Raman--scattered line of HeII ($\lambda$1025.25~\AA) falls at 6545.1~\AA\ (vacuum wavelength; 6543.3~\AA\ air wavelength) and is  observed in PNe such as NGC7027, IC5117, NGC6886, NGC6881  \citep[][and references therein]{choi20}.
Wavelength 6543.3~\AA\ coincides precisely with a DIB in Hobbs'  catalog of HD183143  ($A_c$ of the order of 1.6\%). DIB~6543.3 is therefore likely attributable to IRS of the HeII near--Ly$\beta$ line.
\newpage

\subsubsection{Other lines} \label{oth}
\citet{vangro93} attributed an unknown emission line at 4977~\AA\ in the optical spectrum of symbiotic star RR~Tel to  $\Gamma_{1\rightarrow 2}$ scattering of a CIII transition at 977.03~\AA, close to Ly$\gamma$.
 The exact position of the scattered central wavelength should be 4975.4~\AA\ (air wavelength) and may correspond to DIB~4974.93  \citep[DIB no~32 in][]{fan19}.
 
Following Schmid’s identification of the OVI scattered doublet in symbiotic stars,  \citet[][their Table~1]{nuss89} compiled twelve additional ultraviolet transitions, proximate to Ly$\beta$, of SIII, HeII, OI and CII, and  deemed potential candidates for observation as $\Gamma_{1\rightarrow 2}$ scattered lines within the  wings of H$\alpha$.
The Raman scattered OI doublet and  HeII $\lambda$1025.25 were indeed observed (Sections~\ref{o1} and \ref{he}).
The ultraviolet SIII transitions at  wavelengths 1012.76, 1015.502, 105.567~\AA, as given by NIST, $\Gamma_{1\rightarrow 2}$ scattered at 6065.98, 6165.61, 6168.01~\AA\ (air wavelengths) have  not been reported in symbiotic stars or PNe but  could match DIBs~6065.32, 6165.48, and 6168.06~\AA.
The Raman scattered CII lines  have not been observed and lack DIB counterparts.
\subsection{Wavelength shift of HI Raman scattered ultraviolet emission lines} \label{glin}
The central wavelengths of emission lines are not preserved in scattering processes. 
Thus the observed wavelength discrepancy of 1--2~\AA\  between observed central wavelengths of scattered emission lines and source emission lines' scattered central wavelengths  has not brought into question the Raman origin of the optical emission lines discussed in Sections~\ref{o6},  \ref{he}, and \ref{oth}. 
This discrepancy translates to less than 0.1 \AA\  in the  parent  ultraviolet wavelength space, and was initially attributed to a   relative motion \citep[of the order of 10--20~km/s,][]{schmid89, pequignot97,  lee06} between the scattering HI medium and its ionized envelope.

But the systematic redshift of all observed Raman scattered lines in emission can be explained more  satisfactorily by the dependency of Raman  scattering conversion efficiency on wavelength \citep[][]{pequignot97,  lee06}.  
\citet[][]{jung04} calculated  that  the combination of a line's profile with the Raman cross--section near Ly$\beta$ results in a central wavelength shift of the Raman scattered line exceeding 1~\AA\ in the optically thin limit  ($N_H<10^{20}$~cm$^{-2}$).
This shift decreases as the HI column density increases \citep[as shown in Figure~6 of][]{jung04}, resulting in no wavelength shift in the optically thick limit ($N_H\sim10^{22}$~cm$^{-2}$).
 These calculations, however, ignore dust extinction  \citep[see][]{lee06}, even though dust is the primary cause of ultraviolet interstellar extinction in between Lyman wavelengths.

\begin{figure}
\resizebox{.5\columnwidth}{!}{\includegraphics{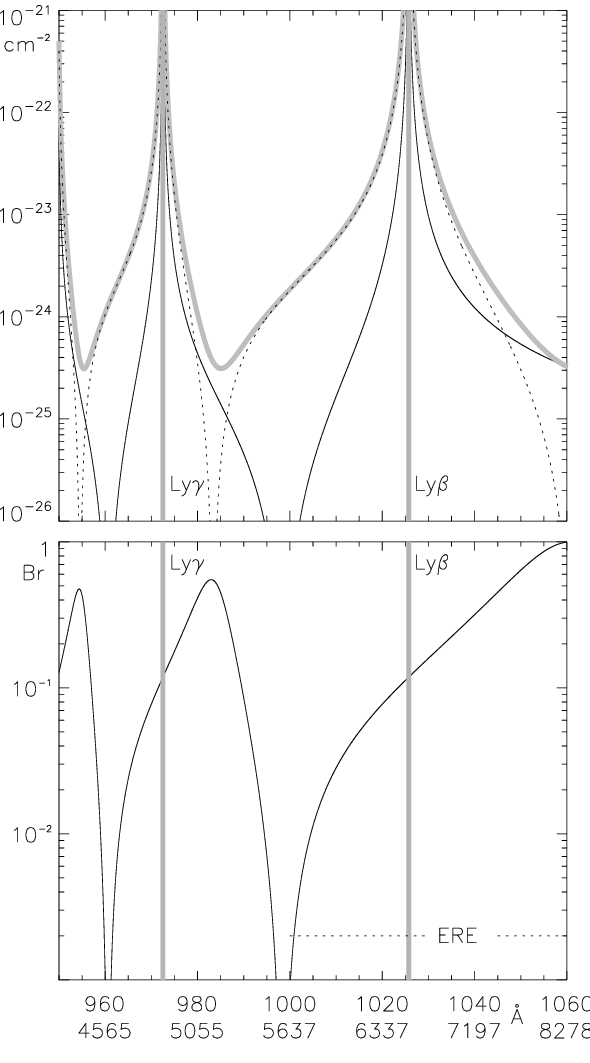}} 
\caption{Bottom plot shows the branching ratio of  $\Gamma_{1\rightarrow 2}$ Raman scattering (data provided by Hee--Won Lee and Seok--Jun Chang) between 950 and 1060~\AA.
This wavelength interval is transformed into the 4000--8000~\AA\ optical wavelength range (bottom line of the x--axis) by  $\Gamma_{1\rightarrow 2}$ scattering.
The ERE wavelength region ranges from $\sim5600$ to 8000~\AA\ ($\sim1000$ to 1050~\AA\ in the parent ultraviolet space).
The top plot shows Rayleigh (dots), $\Gamma_{1\rightarrow 2}$ (plain line), and total hydrogen cross--sections. 
Maxima of $\Gamma_{1\rightarrow 2}$'s branching ratio correspond to minima of the Rayleigh cross--section.\\
Assuming that all HD44179 photons in the vicinity of Ly$\beta$ end up $\Gamma_{1\rightarrow 2}$ Raman scattered in the vicinity of H$\alpha$ or absorbed by dust \citep{apj23}, the proportion of  photons Raman scattered at a given wavelength increases   throughout the  ERE region (bottom plot).
The profiles of  $\Gamma_{1\rightarrow 2}$ scattered ultraviolet emission lines should therefore be characterized by a steep blue rising edge contrasting with a more slowly decreasing red wing.
The scattered lines' central wavelengths will be shifted to the red from the lines' central wavelengths positions.
The red asymmetry should be more manifest for broad lines.
} 
\label{fig:fig2}
\end{figure}

In  \citet{apj23}, I argued that dust should render HI media optically thick to ultraviolet photons at HI column densities much lower than those suggested by \citet[][]{jung04}.
In this case, as shown on Figure~\ref{fig:fig2}, the increase of $\Gamma_{1\rightarrow 2}$'s branching ratio around Lyman wavelengths could explain why Raman scattered ultraviolet lines fall on the red side of their corresponding DIBs and the red asymmetric profiles of DIBs.
Note that these  asymmetric shapes are also reminiscent of Fano profiles, which result from the interaction of discrete energy states  and a continuum, and have been linked to quantum mechanic considerations on Raman scattering \citep{fano61, lounis92}.

\section{Red Rectangle bands (RRBs) and their relation to ERE and DIBs} \label{rrbs}
\subsection{The RRB--DIB issue} \label{ass}
 Figure~1 in \citet[][]{schmidt80} uncovered a few intense, unusually broad, emission bands (RRBs) on top of ERE in the Red Rectangle nebula.
\citet{fossey91, sarre91}  noted a congruence in wavelength  between RRBs $\lambda\lambda$5799, 5855, 6615 and some of the  deepest DIBs, suggesting transitions from the same particles that would either be seen  in emission (RRBs) or in absorption (DIBs).
Figure~4 in \citet[][]{scarrott92} demonstrated that DIB~5797 exhibits an inverted but comparably asymmetric profile, supporting the association between DIBs and RRBs.

This RRB--DIB association was questioned by \citet{glinski02},  based on their observation   that  the central wavelengths of RRBs remain 1--2~\AA\ redward of their associated DIBs' central wavelengths.
Glinsky \& Anderson argue that their finding excludes an emission/absorption process from the same particles.
Recently,  \citet[][see Section~\ref{geo}, this paper]{lai20} also claimed that DIBs and RRBs originate in different carriers.
\subsection{RRB--DIB relationships} \label{vw}
ESO (European Southern Observatory) spectral observations of the Red Rectangle  by  \citet[][]{vanwinckel02}   increased the number of RRBs from less than a dozen to over forty (their Table~3). 
The paper's Table 4 extends the associations between RRBs and DIBs to nine, all of which demonstrate similar 1--2~\AA\ redshifts between RRBs and their corresponding DIBs.
Among these lines, RRB~6204.5 can be observed together with its associated DIB~6203 on its blue side in the spectrum of NGC7027 \citep[][see their Figure~1, and discussion in Section~\ref{geo}, this paper]{zhang05}.

A comparison of Van Winckel et al.'s Table~3 with recent DIB catalogs reveals  additional RRB--DIB associations.
RRBs~6378.6, 6635.1 in Van Winckel et al.'s Table~3 are on the red side of DIBs $\lambda\lambda$6377, 6633, and can certainly be associated with the DIBs.
RRB~6635.1 (associated  to DIB~6633) matches in emission  the absorption  in Orion that \citet{henney21} attributes to the  HI--Raman scattered OI ultraviolet absorption at 1027.4~\AA\  (Section~\ref{o1}).

Weaker RRBs centered close to 6563.4 and 6661.5~\AA\  could  correspond to the  OI Raman scattered lines at 6564.3, 6663.7~\AA\ (air wavelengths) documented in Henney's Table~AI. 
Comparison  to the DIB catalog of \citet{fan19},  shows that other weak RRBs from Van Winckel et al.'s Table~3, such as  $\lambda\lambda$5889.9, 5895.9, 5912.1, may also be linked to weak DIBs, specifically, DIBs~5888.7, 5893.5, and 5893.5.

The most prominent  RRBs described by \citet[][their Table~4]{vanwinckel02} display the same red--extended asymmetric (also termed "red--degraded") and wide  (FWHM up to 20~\AA)  profiles that \citet[][]{scarrott92} have noted for RRB~5799 and its corresponding DIB~5797. 

\subsection{RRBs, ERE, DIBs, and Raman scattering } \label{rrbraman}
The recognition by \citet{witt20} that no existing interstellar dust model meets the observational constraints of ERE observations  lends support to the alternative, quantitatively justified by  \citet{apj23}, that ERE  results from $\Gamma_{1\rightarrow 2}$ Raman scattering (sRS)  of the continuum around Ly$\beta$.
In the context of Galactic low density HI environments, it also follows from  \citet{kokubo24} that no other physical emission process besides  HI Raman scattering of ultraviolet emission lines can account for the observed  characteristics (widths up to 20~\AA\ and red asymmetry) of RRB  profiles.
Consequently, the profile characteristics of RRBs, coupled with their position atop  ERE, provide  compelling evidence that RRBs stem from HI sRS. 

The wavelength shift argument raised by Glinsky \& Andersson against a relationship between RRBs and DIBs (Section~\ref{ass}) rules out an emission versus absorption process, but does not hold in the context of HI Raman scattering (Section~\ref{glin}).
As the same shift is observed between central wavelengths of all known Raman emission lines and the scattered position of the ultraviolet emission lines' central wavelengths, Glinsky \& Anderson's objection turns out to provide further support for the argument that RRBs are Raman scattered features and that DIBs are their IRS counterpart in the spectra of reddened stars.

I conclude that ERE and RRBs in the Red Rectangle arise from sRS at large scattering angles,  within the neutral gas  exposed to the illumination of HD44179, of  the continuum in the vicinity of Ly$\beta$ and of emission lines (such as OI 1027.4~\AA, Raman scattered at RRB~6635).
DIBs that can be associated with RRBs correspond to IRS of the same lines at a null scattering angle, observed in the spectra of stars located far behind  HI interstellar matter. 
These connections between ERE, RRBs, and DIBs explain why DIB density and ERE have the same profile \citep[][and Section~\ref{ds}]{witt14}, which otherwise cannot be accounted for.

\section{The DIB spectrum} \label{ds}
Figure~\ref{fig:fig1} shows the DIB spectrum of HD183143 as of 1995 \citep[][top plot, between  4000 and 9000~\AA]{herbig95} and 2009  \citep[bottom plot, limited to][]{hobbs09}.
Positions of hydrogen lines are indicated by thick gray lines.
The top plot emphasizes the concentration of DIBs  in the H$\alpha$ region, which also hosts the strongest DIBs.
Significant DIBs (accounting for  over 10\% of the continuum)   beyond this region  tend to cluster around the Balmer and Paschen resonances of hydrogen.

At longer wavelengths ($\lambda>1$~$\mu$m), high resolution ground-based observations are limited by telluric absorption.
Two DIBs culminate near Pa$\gamma$ (10935~\AA), as shown in Figure~6 of  \citet[][]{hamano22}.
 Figure~7 in the same study indicates that the strongest near-infrared DIBs above 1~$\mu$m tend to concentrate around Pa$\gamma$ and Pa$\beta$ (12814~\AA).
The top plots of Figure~\ref{fig:fig1} and of Figure~9 in Hamano et al.'s paper confirm a decrease in average DIB depth with increasing distance from the H$\alpha$ region.
These observations follow the expected behavior of Raman scattering (and IRS) by hydrogen, as described in Section~\ref{srs}.

The bottom plot of Figure~\ref{fig:fig1} shows that increasing the sensitivity of observations does not alter these conclusions.
The spectrum of HD183143 by \citet{hobbs09}  essentially led to an increase in the number of DIBs detected in the H$\alpha$ region.
From these data, \citet[][see his Figure~1]{witt14} demonstrated that the density spectrum of DIBs around H$\alpha$ superimposes well with the  Red Rectangle ERE profile.
This resemblance  between spectra of vastly different natures is a consequence of the asymptotic increase of hydrogen's $\Gamma_{1\rightarrow 2}$ cross-section towards Ly$\beta$.
\section{Nonlinearity of IRS, deep DIBs and the 2200~\AA\ bump} \label{bump}
All strong RRBs have a DIB counterpart, but the reciprocal is not true.
The two deepest DIBs, $\lambda\lambda$6284, 5780,  and the also prominent DIBs~6177, 6270, with $A_c$ over 10\%, have no corresponding RRBs in the Red Rectangle.
This finding may reflect a difference in nature between DIBs with and without an RRB emission counterpart.

Anti-Stokes absorptions or absorptions at combinations of Raman shifts are common features of IRS laboratory experiments (Section~\ref{irs}) and may account for DIBs with no associated RRB.
For instance, I  note that  twice the energy difference between levels $n=2$ and $n=5$ of hydrogen falls at 5.71134~eV,  the exact energy of the 2200~\AA\ bump.
The associated wavelength is 2171~\AA, extremely close to the bump's 2175~\AA\ accepted value.
The bump could therefore be the second IRS harmonic of $\Gamma_{2\rightarrow 5}^i$ for self Raman scattering of H$\gamma$ photons.
\section{The geometrical factor in the observation of HI interstellar clouds} \label{geo}
A corollary of  attributing ERE and RRBs to sRS and DIBs to IRS is that ERE and RRBs occur under conditions that exclude the observation of DIBs, and vice versa.
Their respective conditions are determined by the relative alignment of the source radiation field and the line of sight, as well as  the proximity of the source of light to the HI scattering medium.
ERE and Raman scattered emission lines, including RRBs, require intense radiation fields.
They are observed in the vicinity of hot stars, such as symbiotic stars and illuminating stars of nebulae, which implies  large angles of scattering.
In contrast, DIBs are detected in the direction of stars located in the far background of interstellar clouds and do not manifest  in circumstellar matter  (Section~\ref{dibs}).
Observation of ERE, Raman scattered lines, and DIBs from the same direction implies either that different radiation fields are acting on the same medium (see Section~\ref{o1}), or that different HI media are being observed together along the line of sight.
The geometric configuration of the observation thus introduces a critical parameter for interpreting spectra of HI interstellar matter.
The two following examples underscore the importance of  the geometry of an observation in  analyzing HI media spectral data.
\subsection{DIBs and Raman scattered emission lines in NGC7027} \label{ngc}
Weak DIBs in the optical spectrum of NGC7027 \citep{zhang05} have been demonstrated to originate not from the PN itself but from foreground  interstellar matter \citep[][]{lebertre93}.
Conversely, the planetary nebula spectrum exhibits emission features such as ERE \citep{furton90}, the $\Gamma_{1\rightarrow 2}$ scattered HeII   and  OVI  lines \citep{pequignot97,pequignot03,zhang05}, and, just redward of DIB~6203.6, RRB~6206 \citep[Figure~1 in][]{zhang05}, all of which can be attributed to NGC7027's local circumstellar matter.

\subsection{ERE and DIBs in nebula IC63} \label{ic}
\citet{witt90} detected ERE within nebula IC63, which also stands out as the sole nebula where a background star (spectral type A7V) was observed behind the ERE region \citep[][]{lai20}.
The star's spectrum revealed the presence of major DIBs  $\lambda\lambda$4428, 5780, 5797, 6614, leading \citet[][]{lai20} to conclude that ERE and DIBs share common or closely related carriers that differ from those responsible for RRBs', which did not appear in the observations of either Witt \& Boroson or Lai et al.

The conclusions of Lai et al.  do not address the general absence of DIBs in  nebular observations and of ERE in the  spectra of stars.
Since  Lai et al.  acknowledged that "DIBs are usually observed in the line of sight towards a distant star, while ERE is observed most easily in a diffuse emission environment adjacent to a hot illuminating star", the conclusion should be  that detection  of DIBs in IC63 relies entirely on the presence of the A7V background star on the line of sight.
DIBs will be present in the star's spectrum wherever the star is positioned in the backdrop of IC63's ERE field.
Conversely, there should be no  DIBs if there is no background star on the line of sight.
In that case the observation is dominated by IC63 central star's reflected light and leads to the detection of ERE \citep{witt90}.
For reasons that remain to be elucidated, RRBs are known to be weak, blurred, or lacking in nebulae other than the Red Rectangle.
\section{Summary and discussion} \label{dis}
This paper has shown that the observed properties and interrelationships of  three major spectral figures of HI interstellar matter  --extended red emission (ERE), the Red Rectangle bands (RRBs), and the diffuse interstellar bands (DIBs)-- correspond with the differences and  common properties of  two Raman processes: spontaneous Raman scattering (sRS) and inverse Raman scattering (IRS) by atomic hydrogen.
It follows that sRS accounts for ERE and RRBs, and IRS for the DIBs.

sRS of ultraviolet emission lines by atomic hydrogen in its ground state is a well established interstellar phenomenon in symbiotic stars and nebulae.
The scattering is isotropic, and its observation is favored by intense ultraviolet--dominated radiation fields (Section~\ref{srs}).
Consequently, HI sRS of ultraviolet emission lines occurs near bright stars and involves wide scattering angles.
Outputs of HI sRS are primarily observed near the Balmer lines, especially in the H$\alpha$ region (approximately from 5500 to 8000~\AA), which coincides with the ERE wavelength domain.
A particularity of all observed interstellar HI Raman scattered emission lines is a slight redshift  relative to the source emission line central wavelength scattered position on the spectrum (Section~\ref{glin}).

In contrast to near--isotropic sRS, IRS of an emission line is observed exclusively in the direction of the source radiation field, i.e., at a zero scattering angle (Section~\ref{irs}).
Unlike sRS, IRS is a nonlinear stimulated process.
The stimulus is provided by a coherent local continuum encompassing the output frequency,  and has the same direction as the source radiation field.
Instead of producing an emission line as seen in other stimulated Raman scattering (SRS and CARS) or sRS experiments, IRS results in an absorption line within the continuum, centered precisely at the wavelength corresponding to the source line's central scattered position.
Conditions for IRS are  naturally met in observations of stars situated in the distant background of HI interstellar matter (Section~\ref{irsast}).
It can therefore be anticipated that HI Raman scattered emission lines of symbiotic stars and nebulae will appear as absorption features in the spectra of reddened stars, unless  the reddening is caused by local interstellar matter.

Section~\ref{dibid} showed that HI Raman scattered emission lines in the ERE wavelength range observed in symbiotic stars and  nebulae are all represented in DIB catalogs at the exact positions of the Raman scattered source emission lines' central wavelengths  (Section~\ref{dibid}).  
The identification of two oxygen doublets among these lines holds particular significance, given the low probability of coincidental alignment.
As discussed in Section~\ref{srs}, the structure  of the DIB spectrum, which is closely linked to hydrogen’s optical transitions and is largely concentrated around H$\alpha$ (Section~\ref{ds}), and the similarity between the DIB density spectrum and  the ERE profile noted by  \citet{witt14} are consistent with the expectations for HI Raman scattering.

HI sRS of the continuum close to Ly$\beta$  justifies the strength, shape, and wavelength extent of ERE \citep{apj23}.
The widths and red asymmetric profiles of several RRBs, found on top of ERE, point to a Raman origin \citep[][and Section~\ref{rrbraman}]{kokubo24}.
All prominent RRBs are associated with some of the deepest DIBs, with the same wavelength gap observed between known sRS lines and their corresponding DIBs.
RRB~6635.1, associated with DIB~6633.3, should be  $\Gamma_{1\rightarrow 2}$ Raman scattering  of the OI line at 1027.43~\AA\  (Sections~\ref{o1} and \ref{vw}).
I thus concluded that ERE arises from sRS by hydrogen of the continuum near Ly$\beta$, and that RRBs are HI sRS  of emission lines, e.g. close to  Ly$\beta$.
DIBs that can be associated with RRBs  correspond to their IRS counterpart in the spectrum of reddened stars.
Deep DIBs  lacking such an RRB correspondence, and perhaps also the  2200~\AA\ bump of ultraviolet extinction curves, could originate from anti-Stokes IRS or IRS harmonics (Sections~\ref{irs} and \ref{bump}).

The common Raman origin of IRS and sRS explains the large widths and red asymmetry of DIBs and RRBs profiles, the structure of the DIB spectrum, and the similarity of ERE and DIB density  spectra.
Conversely, the marked differences between IRS and sRS  explain why these interstellar spectral features arise from the same interstellar media but are not observed  under the same circumstances  (Section~\ref{geo}).
ERE and RRBs appear in the reflected light from nebulae illuminated by bright stars but are too weak to be observed in the spectra of stars.
DIBs only appear in the complete forward IRS  light of a far--away star, obscured by foreground HI gas, on the line of sight. 
They are notably absent from circumstellar matter.
Observing ERE or sRS HI scattering (including RRBs) and DIBs  in the same direction indicates either that several media are present along the same line of sight (Section~\ref{ngc}) or that several radiation fields impinge on the same medium  (Sections~\ref{o1} and \ref{ic}).
Understanding the geometrical configurations of observations --specifically, the spatial arrangement between the light source, the scattering medium, and the observer-- is therefore critical for a comprehensive interpretation of spectral observations of interstellar matter.

This study confirms that the optical and near--infrared spectral manifestations resulting from the interaction between starlight and photodissociation fronts, including the unidentified infrared bands, ERE, RRBs, blue luminescence, and DIBs, are intricately tied to the spectrum of atomic hydrogen \citep{apj23}. 
These features result from the interaction  of starlight with atomic hydrogen and the elementary atoms and molecules known to be mixed with hydrogen.
This conclusion contrasts sharply with the commonly accepted idea that  ERE, RRBs, and DIBs, although exclusively produced by HI media, are unrelated to hydrogen and instead proceed from the chemistry of complex carbonaceous compounds\footnote{Such as PAHs, Maons,..., see \citet{kwok23}. Note that interstellar dust models proposed by Kwok and others, which are based on the premise that observation of the interstellar optical features depends solely on the presence of these complex molecules along the line of sight,  fail to account for the intensity of ERE \citep{witt20}, negate the RRB--DIB relationship \citep{glinski02,lai20}, and do not explain the attenuation of DIBs when interstellar matter is near a star.
Since the synthesis of these  molecules  cannot occur  in PNe \citep{ar22}, as hypothesized --without justification-- by   \citet{kwok23},
where and how they were  formed remains elusive. 
Why they manifest in low density HI clouds and not in molecular gas is another puzzling question.}.\\

Acknowledgements: I benefited from Alexander Kramida's advice on the use of the National Institute of Standards and Technology (NIST) database. 
Figure~2 is based on cross--sections and branching ratios calculated by Hee--Won Lee and Seok--Jun Chang. 
Air to vacuum wavelength transformations were done with IDL routines provided by  Nikolai Piskunov. 
Jane F. Bestor offered helpful editorial suggestions.
\bibliographystyle{model3-num-names}
{}
\end{document}